\documentclass{article}
\usepackage{spconf,amsmath,graphicx}
\usepackage{multirow}
\usepackage{comment}
\usepackage{hyperref}
\newcommand{\ie}{{\it i.e.}}
\newcommand{\eg}{{\it e.g.}}

\newcommand{\eat}[1]{}
\usepackage[marginal]{footmisc}

\title{Self-supervised learning for audio-visual speaker diarization}
\name{$^1$Yifan Ding, $^2$Yong Xu, $^2$Shi-Xiong Zhang, $^3$Yahuan Cong and $^1$Liqiang Wang\thanks{This works was done while Yifan Ding and Yahuan Cong were interns in Tencent AI Lab, Bellevue, USA.}
}
\address{$^1$University of Central Florida, USA, $^2$Tencent AI Lab, USA, $^3$BUPT, China}
\begin{document}
%
\maketitle
\begin{abstract}

Speaker diarization, which is to find the speech segments of specific speakers, has been widely used in human-centered applications such as video conferences or human-computer interaction systems. In this paper, we propose a self-supervised audio-video synchronization learning method to address the problem of speaker diarization without massive labeling effort. We improve the previous approaches by introducing two new loss functions: the dynamic triplet loss and the multinomial loss. We test them on  a real-world human-computer interaction system and the results show our best model yields a remarkable gain of +8\% {\em $F_1$-scores} as well as diarization error rate reduction. 
Finally, we introduce a new large scale audio-video corpus designed to fill the vacancy of audio-video dataset in Chinese. 

\end{abstract}
\section{Introduction}
\label{sec:intro}



Speaker diarization is the process of partitioning an input audio or video stream into individual segments to match specific speakers. It is one of the core components for many human-centered applications such as video conference systems, human-robot or human-computer interactions, and video re-targeting problems~\cite{howard2017mobilenets,senoussaoui2013study,wang2018speaker}. 
For example, in a human-computer interaction system, multiple people may talk to the system simultaneously, and we need to identify individual active speakers and separate their faces/bodies and audios before analyzing their activities. 

The diarization can be performed on video, audio, or both. Many studies focus on either video-only or audio-only approaches. Everingham et al.  use the movement of lips (\ie, video only) to define active speakers~\cite{Everingham06a}. While in most cases, only audio is used~\cite{senoussaoui2013study,wang2018speaker,zhang2019fully,garcia2017speaker}. Recently, multi-modal (\eg, audio-video) approaches are attracting more attention~\cite{hoover2017putting,marcheret2015detecting,gebru2017audio}. Supervised approaches have been proposed to identify speakers based on the correlation between the audio and video features~\cite{hu2015deep,ren2016look,roth2019ava}, which requires per frame labeling. 
To relieve people from massive labeling work, unsupervised or self-supervised methods are proposed~\cite{shum2013unsupervised,sell2014speaker}. Chung et al. suggest an end-to-end audio-video synchronisation system to use synchronisation between video and audio as the supervision signal. In the approach, a 3D convolutional network is designed to extract video features and contrastive loss is applied~\cite{Chung16a}. 
Recently, Nagrani presents a method for speaker identification/verification~\cite{nagrani2017voxceleb}. In 2018, Korbar et al. presents a self-supervised method is proposed to use curriculum learning and delicate negative example selection strategy, where contrastive loss is employed and claimed as a critical component~\cite{korbar2018co}. 

However, a disadvantage of contrastive loss is that all unsynchronized pairs are treated  equally as negative pairs. Specifically, an audio-video pair with only one frame shift from each other is treated the same as a heterologous pair where the audio and video are from different sources, causing a serious imbalance between positive and negative samples. Besides, human can barely detect lip-sync errors below 200 ms and the training videos downloaded online are sometimes unsynchronized for a few frames due to recording or uploading errors. It harms the model performance to treat these slightly unsynchronized audio-video pairs equally with largely shifted pairs or heterologous audio/video pairs. To relief this problem, Chung et al. tried the classification (\ie, cross entropy) loss but was unable to achieve convergence~\cite{Chung16a}. 

In this paper, we first
propose to sample three kinds of audio-video pairs for training: synchronized pairs, shifted pairs in which audio and video 
are shifted by $j$ video frames, and heterologous pairs where audio and video belong to different sources. 
Then it comes to our major contributions - to propose two new loss functions:  dynamic triplet loss  and multinomial loss. Like standard triplet loss, 
the distance between negative pairs should exceeds the positive pairs by a margin. The difference is that in dynamic triplet loss, positive pairs and negatives pairs are dynamically determined in each iteration. Our experiment shows that it achieves better performance and converges faster than contrastive loss.
But it is still slow because in each iteration, only one positive pair and one negative pair are sampled. It takes many iterations to sample all shifting combinations of each audio-video segment, which makes it harder to find the global optimum.

To solve the above problems, we further propose the multinomial loss, where all shifting combinations for an audio-video pair and all heterologous combinations for audio-video pairs within a mini-batch are considered simultaneously. Specifically, 
we cluster the negative pairs into groups, where different margins with LogSumExp (LSE) are employed to achieve a smooth maximum~\cite{nielsen2016guaranteed} in each group. The experiment results show that multinomial loss achieves even faster convergence and better performance compared with dynamic triplet loss and contrastive loss.


Finally, we propose 
a new large scale audio-video corpus in Chinese~\cite{ourdataset} to fill the vacancy of such kind of training data. 
All our experiments are tested on a real-world human-computer interaction system and the results show the effectiveness of our proposed method.

\section{Our method}
\label{sec:main}






To process multi-modal signals, we use a two-stream network to extract audio and video features separately. The proposed dynamic triplet loss and multinomial loss are employed to optimize the network. 

\subsection{Two-stream network structure}

\begin{figure}[t]
\begin{center}
\includegraphics[width=0.45\textwidth]{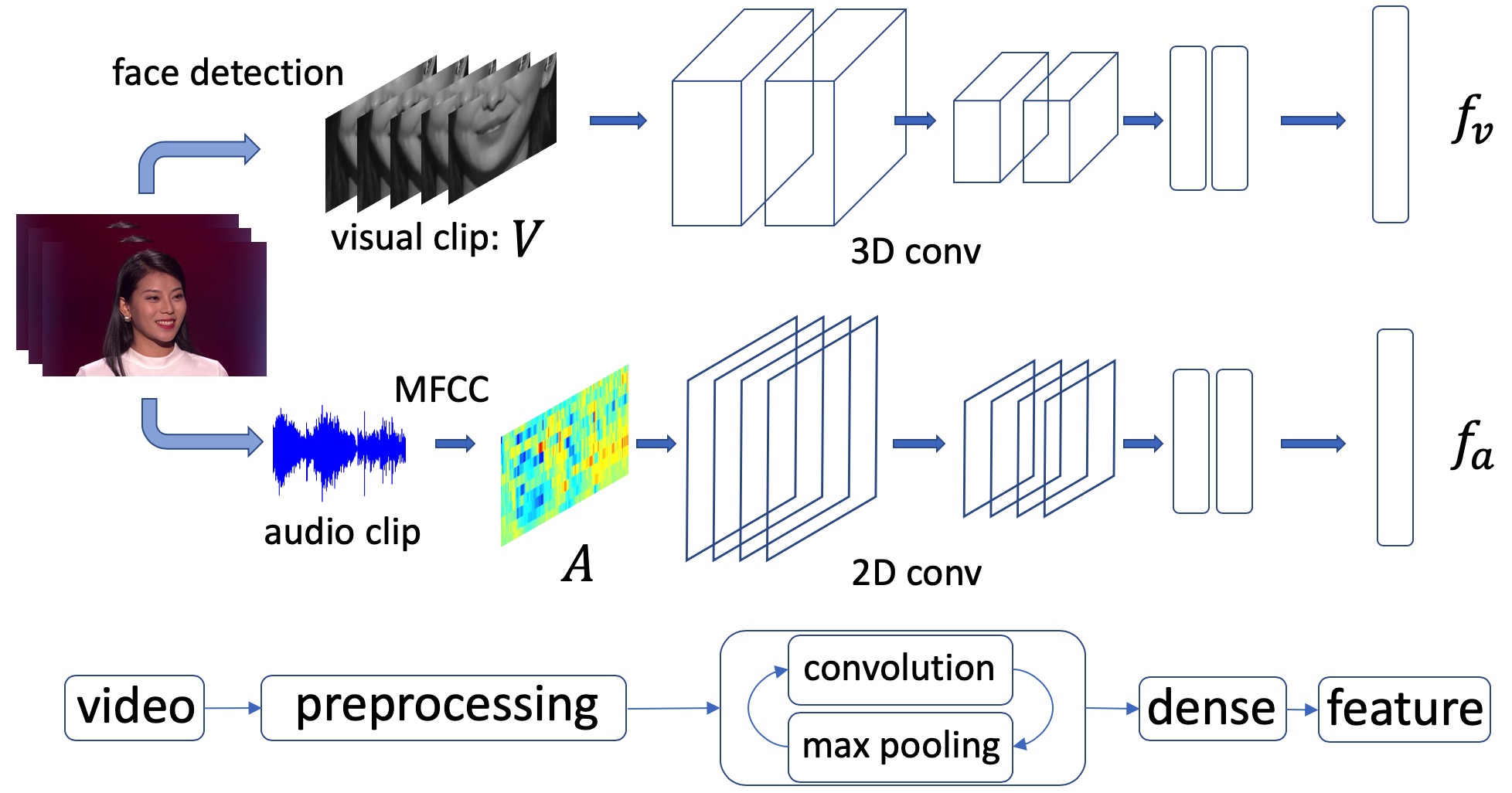}
\end{center}
   \caption{Two stream network architecture.}
\label{fig:network}
\end{figure}

To achieve speech diarizaiton, we process audio and video separately~\cite{Chung16a}, which is a common approach for multi-modal tasks. For audio stream, the input is first transformed to MFCC (Mel Frequency Cepstral Coefficient) features, \ie, a power spectrum of a sound on a non-linear mel scale of frequency. Then the MFCC is sent to a 2D convolutional network to produce speech feature. For video stream, a 3D convolution module is employed to extract both temporal information between consecutive video frames and spatial information in each video frame. 
We use $f_a$ and $f_v$ to denote the audio and video streams, respectively. Figure~\ref{fig:network} shows the structure.

\subsection{Sampling strategy}
\label{subsec:SamplingStrategy}

\begin{figure}[t]
\begin{center}
\includegraphics[width=0.45\textwidth]{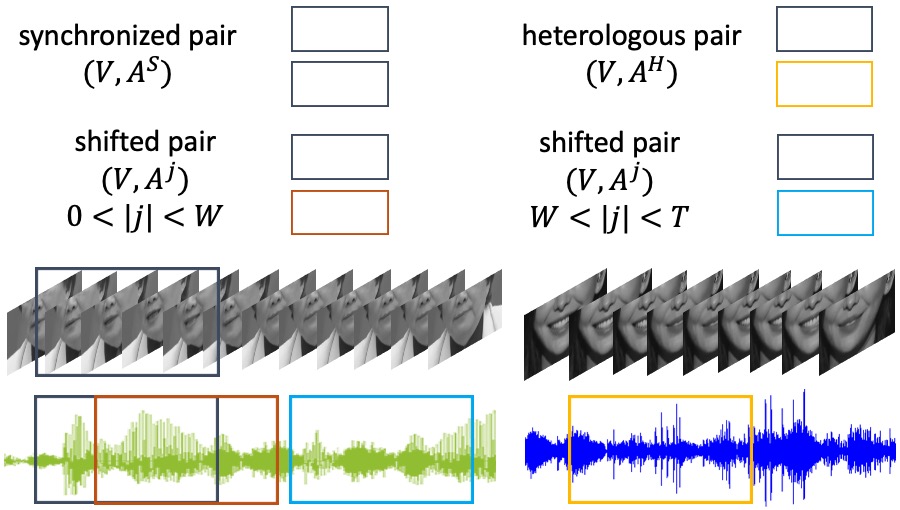}
\end{center}
   \caption{Examples of audio-video synchronized, shifted, and heterologous pairs. $W$ denotes the lenght of visual clip. $T$ is the shifting range.}
\label{fig:cluster}
\end{figure}

Suppose we have a visual segment $V_n$, where $n \in \{1,2...N\}$, $N$ is the total number of visual segments. Then we define its correspondent synchronized audio segment as $A^S_n$.  A shifted audio segment from the same video but with $j$ shifted frame is denoted by $A^j_n$, where $j \in \{-T,...-1,1...T-1,T\}$, $T$ is the pre-defined shifting range, which is 10 in our experiment. Let $A^H_n$ denote a heterologous audio segment from another video. Specifically, we consider three types of audio-video pairs: synchronized pairs $(V,A^S)$, shifted pairs $(V,A^j)$,  and heterologous pairs $(V,A^H)$. All visual and audio segments have consistent length (5 video frames in our experiment). 
The sampled pairs are demonstrated in Figure~\ref{fig:cluster}.  
\begin{figure*}[t]
\begin{center}
\includegraphics[width=0.95\textwidth]{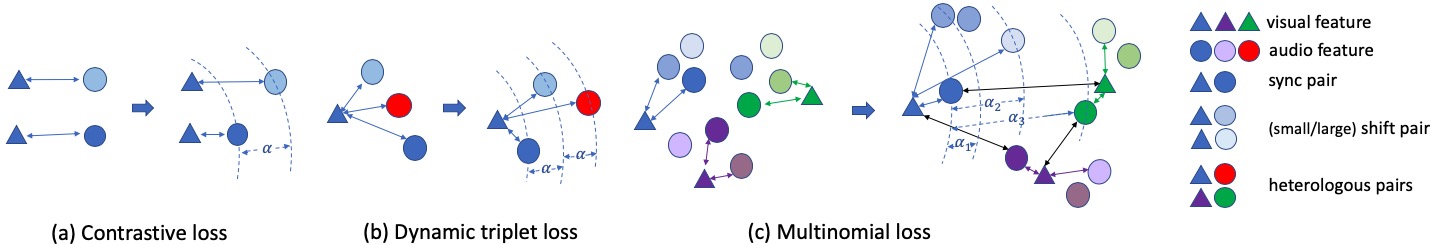}
\end{center}
   \caption{Comparison of loss functions. \textbf{(a):} The negative pair is separated by a margin. \textbf{(b):} Negative pairs are separated with each other by a margin. \textbf{(c):} Negative pairs with different shifts or heterologous audio are separated by separate margins.
   }
\label{fig:overview}
\end{figure*}
\subsection{Dynamic triplet loss}

In contrastive loss, the distance between unsynchronized audio-video pairs are pushed larger than synchronized audio-video pairs. The loss is as follows:
\begin{align}
  L_{con} =  \frac{1}{2N}\sum_{n=1}^N{(y_n)d^2_n + (1-y_n)\max (\alpha - d_n, 0)^2} \\
  d_n = ||f_v(V_n) - f_a(A_n)||_2\label{eq:contrastive-loss}
\end{align}
where $y \in [0,1]$ is the binary similarity metric denoting whether the visual and audio segments are synchronized, and $y = 1$ means synchronized.

A problem in the contrastive loss is that it equally treats all unsychronized pairs. Since negative pairs are sampled by shifting the audio/visual segment or replacing the audio/visual clips from another video. Serious imbalance lies between the number of positive and negative pairs. In other words, there are much more shifted and heterologous pairs than positive synchronized pairs. In addition, the model would more likely be dominated by easy negatives (\eg $(V_n,A_n^H)$ from different videos) rather than hard negatives (\eg $(V_n,A_n^j)$ with small shifts). Finally, the training samples downloaded online could be slightly unsynchronized due to possible recording/uploading error, which may also cause the algorithm to diverge.


In this paper, we propose dynamic triplet loss together with ,.the aforementioned sampling strategy to solve these problems. Specifically, we first sample audio-video training data using the three sampling methods introduced in Section \ref{subsec:SamplingStrategy} to obtain synchronized pairs $(V_n,A_n^S)$, shifted pairs $(V_n,A_n^j)$, and heterologous pairs $(V_n,A_n^H)$. During the training, the positive and negative pairs are dynamically defined according to their relative distance, as shown in Table~\ref{tab:pos-neg}, \ie, a positive pair in one iteration could be a negative pair in another iteration. Finally, the model is optimized with Equation~\ref{eq:triplet-loss}. 
\begin{equation}\label{eq:triplet-loss}
\begin{split}
   L_{D\_tri} = &\sum_{n=1}^N[ ||f_v(V_n)-f_a(A_n^{\prime})||_2^2 \\
  &-||f_v(V_n)-f_a(A_n^{\prime\prime})||_2^2 + \alpha ]_+ 
\end{split}
\end{equation}
where 
$(V,A^{\prime})$ refers to a positive pair, $(V,A^{\prime\prime})$ refers to a negative pair, and $\alpha$ is the pre-defined margin. By using the dynamic triplet loss, the algorithm progressively learns the following distance rule: 
$D(f_v(V_n),f_a(A_n^S)) < D(f_v(V_n),f_a(A_n^{j^\prime})) < 
D(f_v(V_n),f_a(A_n^{j^\prime\prime})) < D(f_v(V_n),\\f_a(A_n^H))$
, where $j^{\prime}, j^{\prime\prime} \in \{-T,...-1,1...T-1,T\}$ and $|j^{\prime}| < |j^{\prime\prime}|$. $D$ denotes the distance function measured by $l_2$. $f_a$ and $f_v$ denote the audio and visual feature extraction functions, respectively.
\begin{table}[h]
\begin{center}
\caption{Definition of positive and negative pairs according to their relative distance.}\label{tab:pos-neg}
\vspace{5pt}
\begin{tabular}{ lcc}
\hline
    
      Index &Positives $(V_n,A_n^{\prime})$ & Negatives $(V_n,A_n^{\prime\prime})$\\
\hline\hline
Case-1 &$(V,A^S)$ & $(V,A^j)$ \\
Case-2 &$(V,A^{j^{\prime}})$ & $(V,A^{j^{\prime\prime}}) (|j^{\prime}| < |j^{\prime\prime}|)$ \\
Case-3 &$(V,A^{j})$ & $(V,A^H) $ \\
\hline
\end{tabular}
\end{center}
\end{table}


\subsection{Multinomial loss}

In dynamic triplet loss, only two pairs are sampled in each iteration. However, for each audio-video pair, there are $\mathrm{C}_{2T}^2$ shifting options and more heterologous options. To best take advantage of these combinations during each training iteration, we further propose an improved loss called {\it multinomial loss} by optimizing the positive pairs and negative pairs by clusters, while each cluster has its own optimization margin. In this way, the negative pairs with different specialities (\eg shifting range) would be treated differently. We propose to cluster the negative pairs into groups and apply separate margins for each group. Equation~\ref{eq:cluster-loss1} defines the loss.

\begin{equation}\label{eq:cluster-loss1}
\begin{split}
  L_{mul}& =   \sum_{n=1}^{N}( D(f_v(V_n),f_a(A_n^S))  \\
  & +\sum_k^{K}{\log(\sum_u^{u \in cluster\{k\}}{\exp(\alpha_k - D(f_v(V_n),f_a(A_n^u))}) }  )
\end{split}
\end{equation}

\begin{equation}\label{eq:cluster-loss}
\begin{split}
  L_{mul}& =   \sum_{n=1}^{N}( D(f_v(V_n),f_a(A_n^S))  \\
  & +\log(\sum_{j^\prime}^{0 < |j^{\prime}| \leq m_1}{\exp(\alpha_1 - D(f_v(V_n),f_a(A_n^{j\prime}))}) \\
  & +\log(\sum_{j^{\prime\prime}}^{ m_1 < |j^{\prime\prime}|\leq m_2}{\exp(\alpha_2 - D(f_v(V_n),f_a(A_n^{j\prime\prime}))})\\
  & +\log(\sum_H^{H\neq n}{\exp(\alpha_3 - D(f_v(V_n),f_a(A_n^H))})   )
\end{split}
\end{equation}

Specifically, in our audio-video synchronization learning method, we separate our loss into four parts, which is shown as Equation~\ref{eq:cluster-loss}. The first part is to minimize the distance of the synchronized pairs $D(f_v(V_n),$ $f_a(A_n^S))$. The second part denotes the loss for shifted pairs when the shifting distance is within $m_1$. We set $m_1$ to be equal to the size of video segment, which is 5 frames in our experiment
. {\em LogSumExp}~\cite{nielsen2016guaranteed} is applied to achieve a smooth maximum and $\alpha_1$ is a margin to this loss. The third and fourth loss functions are similar but for different audio-video pairs. The third loss is for shifted audio-video pairs where the shifting range from $m_1$ to $m_2$, when the shifted audio and visual segments are from the same video but not temporally overlapped. We set $m_2$ to be 10 frames and $\alpha_2$ is the corresponding margin. The forth loss and margin $\alpha_3$ is for heterologous pairs, \ie, visual and audio segments from different videos in the mini-batch.



\section{Experiments}
\label{sec:exps}

In this section, we describe the dataset we uses, the test metrics and the experimental results.

\subsection{Dataset and baselines}

Our training dataset contains over 140 thousand video clips, about 100 hours long in total. All video clips are from YouTube with the front face facing the camera. The test-set includes 12 videos. Each is 40-60s long and contains 3 to 4 people talking in turn. Our data would be described in details and released soon in another paper~\cite{ourdataset}. 
We compare our results with SyncNet~\cite{Chung16a} and UIS-RNN (unbounded interleaved-state recurrent neural networks)~\cite{zhang2019fully}. UIS-RNN is a fully-supervised audio-only speaker diarization system which takes d-vector embedding as input and each individual speaker is modeled by a parameter-sharing RNN, while the RNN states for different speakers interleave in the time domain. We also extended the UIS-RNN \cite{zhang2019fully} with the number of detected faces as the interleaved-state upbound in RNN. This means the number of faces in the video are used to indicate the number of potential speakers.





\subsection{Testing results}
\vspace{-10pt}
\begin{table}[th]
\begin{center}
\caption{Model Comparison (\%))}\label{tab:results}
\vspace{2pt}
\begin{tabular}{ ll|cc}
\hline
    
      \# &Method& {$F_1$-scores} & {\em DER }\\
      \hline
      1& SyncNet~\cite{Chung16a}&76.4& 23.1\\
      2& UIS-RNN~\cite{zhang2019fully}&70.0&25.6\\
      3& face-bounded UIS-RNN &72.6&22.2\\
       
      4& ours-triplet&75.3&22.9\\
      5& ours-triplet-LipNet\cite{chung2017lip}&76.8&21.7\\
      6& ours-cluster&\textbf{84.9}&\textbf{17.0}\\
\hline
\end{tabular}
\end{center}
\end{table}
In this section, we compare our results with the baselines. For all the models, the faces are detected with Dlib~\cite{dlib}. Gray-scale images of the lower half face is resized to be $112\times 112$ for training. For the audio part, a 13 mel MFCC feature are used in our methods. We use 25FPS for the visual clips and 100Hz for the audio segments. Therefore the length of visual and audio input is 5-frame and 20-frame respectively. The value of $\alpha_1,\alpha_2 and \alpha_3$ are set to be 1,2 and 10 respectively. 
A batch-size of 16 is used for all experiments and no data augmentation is implemented. Besides, the distances are generated through a per-frame evaluation of the $l_2$ distance between the audio and visual feature for each speaker in the video. The speaker (\ie face) feature which has the lowest distance with the audio feature would be identified as the active speaker. Table~\ref{tab:results} shows the results.

\begin{figure}[h]
\begin{center}
\includegraphics[width=0.45\textwidth]{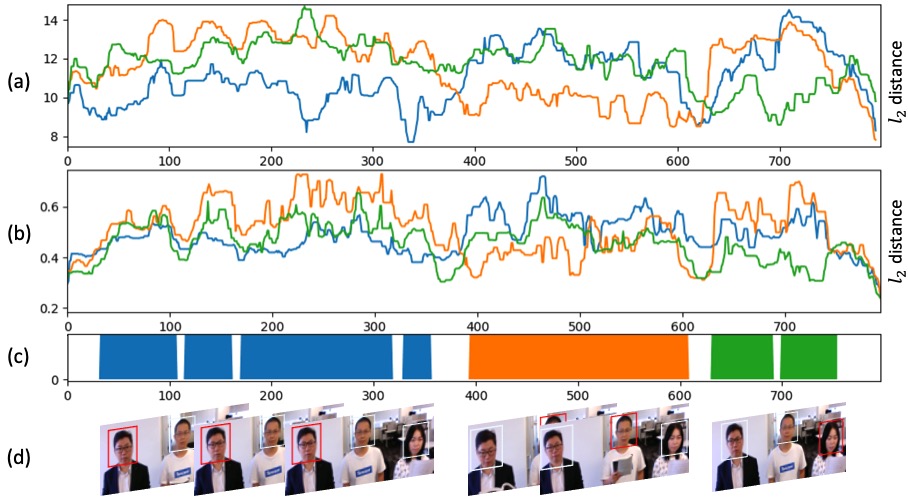}
\end{center}
   \caption{Per-frame test distances. \textbf{(a):} Ours. Different colors denote the a-v distance of different speakers. The curve with lowest distance is the predicted active speaker. \textbf{(b):} SyncNet.  \textbf{(c):} GT.  \textbf{(d):} Visualization of GT. The red box frames the face of active speakers.  }
   \vspace{-10pt}
\label{fig:dist}
\end{figure}

We use {\em DER }(Diarization Error Rate) and {\em $F_1$-scores} to evaluate the performance of each model. 
Compared with SyncNet~\cite{Chung16a}, our best model improved 8\% and 6\% on {\em $F_1$-scores} and {\em DER}, respectively. 
To make it clear, the same MLP network is employed in model \#1 (``SyncNet" ) and model \#6 (``ours-cluster").
However we find incorporating LipNet~\cite{chung2017lip} is also beneficial to the performance since LipNet has a more complex architecture (\ie ResNet~\cite{he2016deep}) compared with 6-layer MLP (multi layer perceptron) we used in model \#4 (``ours-triplet" in Table~\ref{tab:results}).  
Figure~\ref{fig:dist} shows the visualization of per-frame audio-video distances generated by our proposed multinomial model (``(a)") and SyncNet (``(b)"). From which we can find: for our model, the distance of the active speaker is significantly below the distance of non-active speakers. While for SyncNet, the distances of different speaker could hardly be distinguished. 
While in More video demos can be found from the project homepage.\footnote{\href{https://yifan16.github.io/av-spk-diarization}{https://yifan16.github.io/av-spk-diarization}}

\section{Conclusion}
\label{sec:conclu}

In this paper, we propose two new losses: the dynamic triplet loss and multinomial loss, and a large scale dataset in Chinese 
for self-supervised audio-video synchronization learning. The work can benefit the task of speaker diarization, which is an important basic task for many human-centered applications. 
We demonstrate experiments on a real-world human-computer interaction system and compare our results with several baselines. The results show the proposed methods outperforms both the audio-only and multi-modal previous approaches.

\bibliographystyle{IEEEtran}
\bibliography{strings,refs}

\end{document}